\documentclass[pra,onecolumn,floatfix,superscriptaddress,longbibliography,notitlepage, nofootinbib]{revtex4-1}
\pdfoutput=1
\usepackage{tikz}
\usepackage[utf8]{inputenc}
\usepackage{braket}
\usepackage{amsmath}

\usepackage{dcolumn}   
\usepackage{bm}        
\usepackage{amssymb}
\usepackage{mathtools}
\usepackage{tikz}
\usepackage{qcircuit}
\usepackage{appendix}
\usepackage{hyperref}
\usepackage{subcaption}
\usepackage{amsmath,amsfonts,amssymb}
\usepackage{mathtools}
\usepackage{thmtools,thm-restate}
\usepackage{algorithm}

\usepackage{algpseudocode}

\newtheorem{definition}{Definition}
\newtheorem{theorem}{Theorem}
\newtheorem{lemma}{Lemma}

\usetikzlibrary{arrows.meta}

\def\BibTeX{{\rm B\kern-.05em{\sc i\kern-.025em b}\kern-.08em
    T\kern-.1667em\lower.7ex\hbox{E}\kern-.125emX}}

\begin{document}
\title{Quantum Algorithm for Computing Distances Between Subspaces}
\author{Nhat A. Nghiem}
\affiliation{Department of Physics and Astronomy, State University of New York at Stony Brook, Stony Brook, NY 11794-3800, USA}

\begin{abstract}
Geometry and topology have generated impacts far beyond their pure mathematical primitive, providing a solid foundation for many applicable tools. Typically, real-world data are represented as vectors, forming a linear subspace for a given data collection. Computing distances between different subspaces is  generally a computationally challenging problem with both theoretical and applicable consequences, as, for example, the results can be used to classify data from different categories. Fueled by the fast-growing development of quantum algorithms, we consider such problems in the quantum context and provide a quantum algorithm for estimating two kinds of distance: Grassmann distance and ellipsoid distance. Under appropriate assumptions and conditions, the speedup of our quantum algorithm is exponential with respect to both the dimension of the given data and the number of data points. Some extensions regarding estimating different kinds of distance are then discussed as a corollary of our main quantum algorithmic method. 
\end{abstract}
\maketitle

\section{Introduction}
Quantum computation has opened up a completely new frontier in computational science. A vast amount of difficult computational problems have been theoretically shown to be accelerated by quantum computation. Famous examples include integer factorization~\cite{shor1999polynomial}, unstructured database search~\cite{grover1996fast}, quantum simulation~\cite{lloyd1996universal}, the task of probing properties of a black-box function~\cite{deutsch1992rapid}, solving linear system~\cite{harrow2009quantum, childs2017quantum},  and approximating topological invariants~\cite{aharonov2006polynomial}. More recently, the interplay between quantum science and machine learning, so-called quantum machine learning, has led to many fascinating works, such as quantum neural network~\cite{schuld2018supervised, killoran2019continuous}, quantum convolutional neural network~\cite{cong2019quantum}, quantum support vector machine~\cite{rebentrost2014quantum}, etc. Under certain assumptions regarding input access, the exponential speedup is achievable, such as performing supervised learning and unsupervised learning using quantum processors~\cite{lloyd2013quantum} and performing fitting over large data set~\cite{wiebe2012quantum}. Unconditional proof of quantum advantage was provided in~\cite{bravyi2018quantum}, where the authors showed that shallow circuits can completely outperform their classical counterparts. \\

Aside from the aforementioned instances, where the domain of investigation ranges from algebraic problems to data science \& machine learning, the potential advantage of quantum computers has also been explored in (computational) topology \& geometry domain. Lloyd et al.~\cite{lloyd2016quantum} provided a quantum algorithm, the so-called LGZ algorithm, for computing Betti numbers of simplicial complexes, a classic problem arising from topological data analysis. Many followed-up works, such as~\cite{gyurik2022towards, hayakawa2022quantum, mcardle2022streamlined, ubaru2021quantum} have  improved the running time, as well as implementation costs of the LGZ algorithm. In~\cite{ambainis2020quantum}, the authors outlined a quantum algorithm for problems in computational geometry, showing significant speedup. In~\cite{nghiem2022constant},  a single-query, constant-time quantum algorithm is provided for detecting the homology class of closed curves, which is an interesting problem in computational topology. These examples suggest a potentially fruitful domain where quantum advantage could be further explored, since topology and geometry, while being a classical subject within the pure mathematical domain, have provided a solid foundation for many applications. For example, the field of computational conformal geometry~\cite{gu2008computational}, which encompasses modern differential geometry, Riemann geometry, and algebraic topology, has provided many useful tools for challenging problems in computer vision, medical imaging, such as surface classification, registration, etc.  \\

Motivated by these developments, particularly regarding geometry \& topology, we tackle the following problem that arises from the same context: computing distance between linear subspaces \cite{ye2016schubert}. This problem enjoys great applications in machine learning and computer vision, etc., but is quite challenging from a computational point of view since it requires the ability to evaluate singular values of possibly large matrices. We will show that quantum algorithmic techniques could enhance such tasks, yielding significant speedup compared to standard classical methods under certain conditions. Our work thus contributes as another example where quantum algorithms could be beneficial for practical problems. \\

The structure of the paper is as follows. In Section~\ref{sec: classical}, we begin with some introduction to topology and geometry so as to review some preliminaries and build up intuition that is very beneficial to understand the framework behind two problems that we wish to solve, which are computing the distance between \textit{linear subspaces} and between \textit{ellipsoids}. The formal description and the classical solution to those problems are also presented accordingly. In Section~\ref{sec: preli}, we introduce some necessary recipes and tools that are crucial in our subsequent construction. Section~\ref{sec: quantum} is dedicated to our main result, which is an efficient quantum algorithm for computing the linear subspace distance, as well as giving details on error analysis and the final statement regarding its running time. The quantum solution to the second problem---computing the distance between ellipsoids is presented in Section~\ref{sec: quantumellip}, which is more or less an adaptive version of the method outlined in  Section~\ref{sec: quantum} and a result combined with some well-known quantum algorithms, such as inverting a dense matrix~\cite{wossnig2018quantum}. We then show that how the main algorithms presented in section~\ref{sec: extension} can be extended to estimate other kinds of distance between subspaces. In section~\ref{sec: memorygrassman}, we showcase how Grassmann distance and ellipsoid distance are estimated in the so-called memory model, which is a special type of quantum data structure that allows more subtle loading of classical data. We then conclude with some comments on our results in Sec.~\ref{sec:conclusion} and discuss future prospects. 

\section{Distance Between Subspaces}
\label{sec: classical}
Intuitively, geometry and topology typically begin with a \textit{set of points} and other sets built from them, resulting in the so-called space. The most familiar space is probably the Euclidean space of a given dimension $n$, denoted as $R^n$, where each `point' is associated with an $n$-tuple $(x_1,x_2, ..., x_n)$ (also called coordinates), where each $x_i \in R$. The collection of such points, endowed with addition and scalar multiplication operations, forms a \textit{linear vector space}. These concepts belong to a subject called linear algebra, which has found countless applications in science \& engineering, due to its simplicity and versatility. A remarkable property of such Euclidean space is that the \textit{distance} between two points can be defined as some function of their coordinates. The Euclidean space is an instance of a more general concept called \textit{manifold}, with the distance between two points now generalized as the \textit{geodesic distance}. \\

\begin{figure}[htbp]
    \centering
    \begin{tabular}{cc}
        \includegraphics[width = 0.9\textwidth]{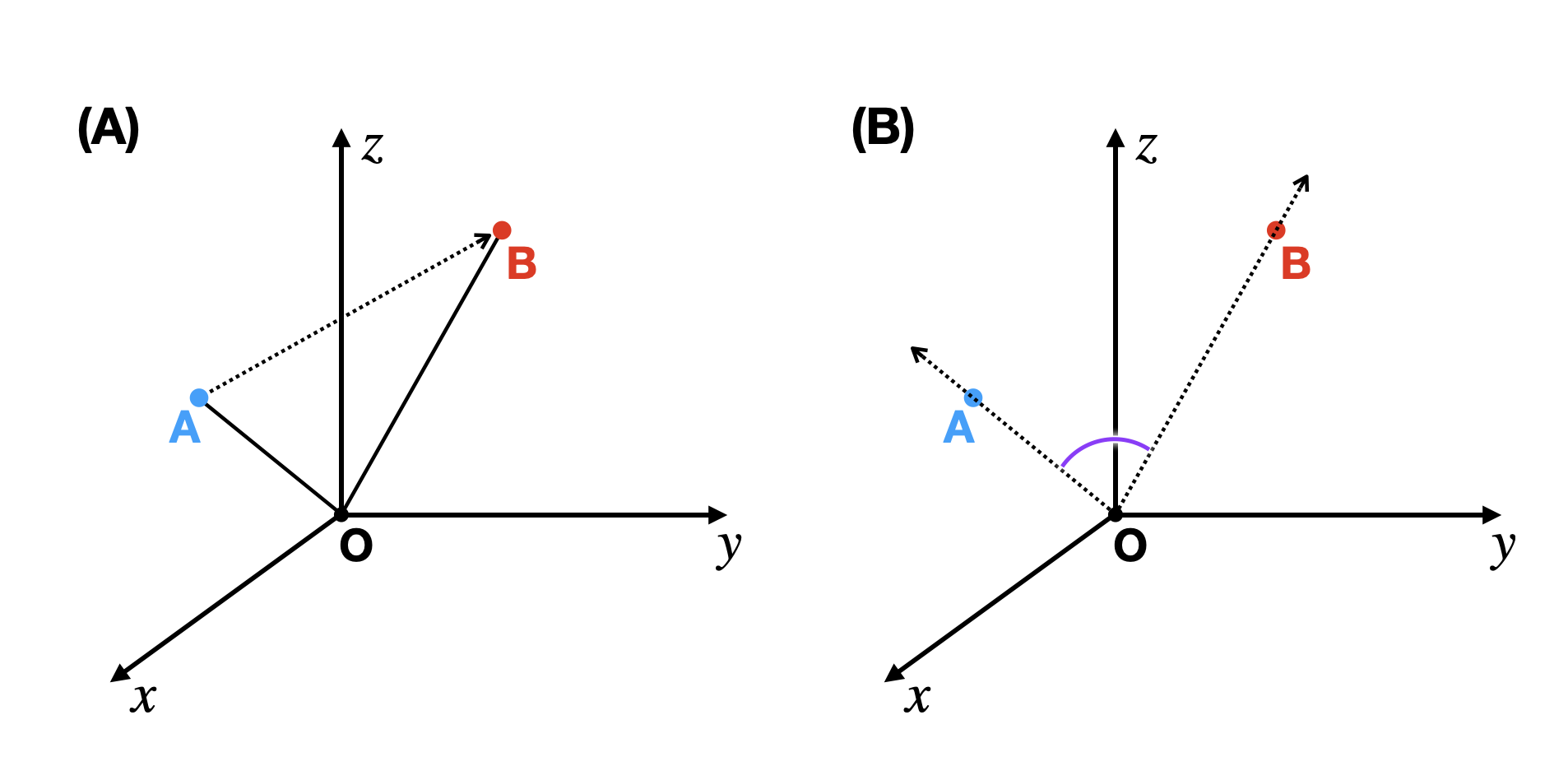}
    \end{tabular}
    \caption{ \textbf{Left figure:} An elementary example of Euclidean space $\mathbb{R}^3$ with two points A and B (and original point O). The distance between A and B can be defined simply based on corresponding coordinates of A and B. \textbf{Right figure:} Illustrative example of distance between two subspaces. Instead of viewing A and B as two points, we consider two 1-dimensional subspaces each spanned by $\Vec{OA}$ and $\Vec{OB}$. Intuitively, the angle between $\Vec{OA}$ and $\Vec{OB}$ can be used as a measure of `distance' between two 1-dimensional subspace.    }
    \label{fig: space}
\end{figure}

\noindent\textbf{\textit{Linear Subspaces:}}  Given a vector space $R^n$, let $\{m_1, m_2,..., m_k\}$ and $\{n_1, n_2, ..., n_k\}$ be two collections of linearly independent vectors. Without loss of generalization, we can assume them to be orthonormal sets. Denote $M_A$ and $M_B$ as two $k$-dimensional vector subspaces spanned by  $\{m_1, m_2,..., m_k\}$ and $\{n_1, n_2, ..., n_k\}$, respectively. Our main problem to compute the separation between $M_A$ and $M_B$. As suggested in~\cite{ye2016schubert}, $M_A$ and $M_B$ could be regarded as two elements, or two points, of the so-called Grassmannian manifold, Gr($k$,$n$). The separation between $M_A$ and $M_B$ could be computed as geodesic distance $d_{M_A,M_B}$ between two points on manifold Gr($k$,$n$). Following~\cite{ye2016schubert}, we now describe the approach for computing $d_{M_A,M_B}$. 

First, we organize $\{m\}: = \{m_1, m_2,..., m_k\}$ and $\{n\} := \{n_1, n_2, ..., n_k\}$ as two column matrices $M,N \in R^{n \times k}$ (which means that the $i$-th column of $M,N$ are $m_i,n_i$ respectively). We now perform singular value decomposition (SVD) of $M^T N$:
\begin{align}
    M^T N = U \cdot \Sigma \cdot V^T,
    \label{eqn: SVD}
\end{align}
with $\sigma \in R^{k\times k}$ having diagonal entries $\{\sigma_i\}_{i=1}^k$ as singular values of $M^T N$. The so-called Grassmann distance between $M_A$ and $M_B$ can be computed as:
\begin{align}
    d_{M_A, M_B} = \sqrt{ \sum_{i=1}^k \theta_i^2 },
\end{align}
with $\cos (\theta_i) = \sigma_i$. We remark that since $\{m\}$ and $\{n\}$ are orthonormal bases, then $0 \leq \sigma_i \leq 1$ for all $i$. \\

It is clear that the computational expense lies mostly in this SVD step. The running time of a classical algorithm that can compute the above Grassmann distance is obviously polynomial in $n$ and $k$. The key steps include multiplying $M^T$ to $N$ and performing SVD on $M^T N$. The multiplication of $k \times n$ matrix $M^T$ and $n \times k$ matrix $N$ takes time $\mathcal{O}( n^2 k)$, meanwhile performing SVD on a $k \times k$ matrix using the so-called Jacobi rotations~\cite{drmavc2008new} takes $\mathcal{O}(k^3)$, yielding the total running time $\mathcal{O}(n^2k + k^3)$. There are several works that provide quantum algorithms for performing SVD, such as \cite{rebentrost2018quantum, bellante2022quantum}. While there is a certain overlap, our subsequent quantum procedure employs somewhat different techniques since the data input to our problem is different, and in particular, our objective is also different.   \\

\noindent\textbf{\textit{Ellipsoids:}} Given a real symmetric positive definite matrix $M \in R^{n \times n}$, an ellipsoid $\mathcal{E}_A$ is defined as 
\begin{align}
    \mathcal{E}_A = \{x \in R^n: x^T M x \leq 1 \}.
\end{align}

The problem is that we need to find the separation, or distance, between two ellipsoids $\mathcal{E}_M$ and $\mathcal{E}_N$. A simple way to compute it is based on the \textbf{metric} defined on the cone of real symmetric positive definite matrices~\cite{ye2016schubert}, which yields distance $\delta_{M,N}$ between $\mathcal{E}_M$ and $\mathcal{E}_N$:
\begin{align}
    \delta_{M,N} = \sqrt{ \sum_{i=1}^n \log^2 (\lambda_i(M^{-1}N)) },
\end{align}
where $\lambda_i$ is the $i$-th eigenvalue of $A^{-1}B$. Since both $A$ and $B$ are real positive definite, $\{\lambda_i\}_{i=1}^n$ are real and positive. The running time of a classical algorithm that computes the above distance is similar to that of Grassmann distance,  which is $\mathcal{O}(n^3)$, as they share similar computational steps, and in this case, we have that $k = n$.

\section{Some Preliminaries}
\label{sec: preli}
The previous section has provided some introduction and classical methods to compute Grassmann and ellipsoid distance. This section provides a review of necessary quantum tools that we would later use to develop our quantum algorithms. 

\begin{definition}[Block Encoding Unitary]
\label{def: blockencode}
Let $A$ be some Hermitian matrix of size $N \times N$. Let a unitary $U$ have the following form:
\begin{align*}
    U = \begin{pmatrix}
       A & \cdot \\
       \cdot & \cdot \\
    \end{pmatrix}.
\end{align*}
Then $U$ is said to be a block encoding of matrix $A$. Equivalently, we can write:
\begin{align*}
    U = \ket{ \bf{0}}\bra{ \bf{0}} \otimes A + \cdots
\end{align*}
\end{definition}
where $\ket{ \bf{0}}$ denotes the first computational basis state in some larger Hilbert space. It is quite obvious from the above definition that 
\begin{align}
    A_{ij} = ( \bra{\bf{0}} \otimes \bra{i} ) U  ( \ket{\bf{0}} \otimes \ket{j}  ),
    \label{eqn: rowcolumn}
\end{align}
where $A_{ij}$ refers to the entry of $A$ at $i$-th row and $j$-th column. \\

The next one is a very powerful technique that is often dubbed as quantum signal processing \cite{gilyen2019quantum, low2017optimal}. 
\begin{lemma}[Quantum Signal Processing \cite{gilyen2019quantum}]
\label{lemma: qsp}
Let f be a polynomial of degree d with $|f(x)| \leq 1/2$ for all $x \in [-1,1]$. Let U be a block encoding of some Hermitian matrix A. Then the following transformation
\begin{align*}
     \begin{pmatrix}
       A & \cdot \\
       \cdot & \cdot \\
    \end{pmatrix} \longrightarrow 
     \begin{pmatrix}
       f(A) & \cdot \\
       \cdot & \cdot \\
    \end{pmatrix}
\end{align*}
is the so-called  block encoding of $f(A)$ and can be realized using $d$ applications of $U$ and $U^T$ plus one controlled-$U$ gate.
\end{lemma}

The above Lemma is the most general statement of quantum signal processing technique, which shows flexibility in how the transformation of given matrix could be done. As being pointed out in \cite{low2017optimal}, by making use of Jacobi-Anger expansion, one can efficiently transform the Hamiltonian operator $H$ into a high-precision approximation operator of $\exp(-iHt)$. The scaling cost turned out to be optimal, which demonstrate the surprising power of such quantum signal processing technique. \\

Before proceeding further, we make the following remark. At the core of quantum signal processing technique is the operation on the block encoded matrix, or operator (see Lemma \ref{lemma: qsp}). The transformed operator is apparently block encoded at the same block e,g, top left corner. If we wish to apply this operator to some state and extra the measurement outcome from the output state, how can it be done in the subspace of the encoded operator? The answer is simple. Let $U$ denotes the unitary block encoding of some operator $A$. Let $\ket{\bf 0} \ket{\Phi}$ denotes some state sharing the same dimension as $U$, where $\ket{\Phi}$ has same dimension as $A$, and obviously $\ket{\bf 0}$ accounts for the remaining dimension. As we shall also show later, if we apply $U$ to $\ket{\bf 0}\ket{\Phi}$, then we have:  \\
\begin{align}
    U \ket{\bf 0}\ket{\Phi} = \ket{\bf 0} A\ket{\Phi}  + \sum_{j\neq 0} \ket{j}\ket{Garbage_j}
\end{align}

If we perform measurement on the above state and conditioning on the extra register being $\ket{\bf 0}$ (success with probability $|A\ket{\Phi}|^2$), then the resulting state is $A\ket{\Phi}/|A\ket{\Phi}|$, which is of our interests. If we, for example, perform further measurement on such a state, then the probabability of measuring a basis outcome $\ket{i}$ is $|\braket{i, A\ket{\Phi}}/|A\ket{\Phi}| |^2$.  Therefore, if we conditionally choose the first register to be $\ket{\bf 0}$, then the probability of measuring $\ket{\bf 0}\ket{i}$ is simply $|\braket{i, A\ket{\Phi}}|^2$, which is equivalent to measuring the vector of interested $A\ket{\Phi}$ in computational basis. Therefore, the effect of high-dimension of the encoding step can be trivially removed by simply conditioning on $\ket{\bf 0}$ subspace. Throughout the work, we would simply work in the subspace of encoded operator only, for simplicity. \\

The last recipe we need to construct our quantum algorithm is called an efficient matrix application, from a simple adaptive version of quantum random walk method~\cite{childs2017quantum} and is stated in  the following lemma.

\begin{lemma}[Efficient Matrix Application]
Given coherent oracle access to some s-sparse (the number of maximum non-zero entries in each row or column), Hermitian matrix $H$ of dimension $n\times n$, and a given $n\times 1$ state $\ket{b}$. Then there is a unitary $U_H$ that acts in the following way,
\begin{align*}
 U_H \ket{0^m}\ket{b} = \ket{0^m} (H/s)\ket{b} + \ket{\Phi_{\perp}},
\end{align*}
where $\ket{\Phi_{\perp}}$ 
is some unimportant state (not properly normalized) that is orthogonal to 
$\ket{0^m} (H/s)\ket{b}$, i.e,  
$\ket{0^m}\bra{0^m} \otimes \mathbf{1} \ket{\Phi_{\perp}} = 0$. 
The unitary $U_H$ runs in time 
\begin{align*}
 \mathcal{O}\Big( \log (n) +  \log^{2.5} (\frac{1}{\epsilon}) \Big), 
\end{align*}
where $\epsilon$ is the error tolerance. 
\label{lemma: matrixapp}
\end{lemma}

Since the above prior results were collected for our purpose, we refer the readers to their respective original works for more thorough treatment. Now, we proceed to apply them to construct our main quantum algorithms.  

\section{Quantum Algorithm for Linear Subspaces Distance}
\label{sec: quantum}
This section is dedicated to the first problem that we described in a previous section: computing the Grassmann distance between linear subspaces, each spanned by an orthonormal set. We sketch the main idea behind our quantum algorithm as follows.

First, we remind that the goal is to perform SVD~\ref{eqn: SVD} on $M^T N$ to obtain the singular values for subsequent algebraic operations. In order to achieve such a goal, we need to be able to simulate $\exp( -i M^T N t)$, followed up with a phase estimation procedure (similar to the HHL algorithm~\cite{harrow2009quantum}) to `extract' the singular values and perform our desired calculation from these values.
However, as $M^TN$ is not necessarily symmetric, and hence might not be diagonalizable. We thus  aim to simulate $\exp(-i (M^TN)^T M^T N t )$ instead. The reason is that, while $M^TN$ is not symmetric, $(M^TN)^T M^T N$ is symmetric, and hence being diagonalizable, with the eigenvalues of $(M^TN)^T M^T N $ are the square of the singular values of $M^T N$. Furthermore, as we mentioned in a previous section, the singular values of $M^T N$ are guaranteed to be positive, and hence, we will not suffer from the sign problem if we take the square root of eigenvalues of $(M^TN)^T M^T N$.  \\

The following section is devoted to the  simulating  of $\exp(-i (M^TN)^T M^T N t )$.

\subsection{Encoding Matrix }
\label{sec: preparingstate}
We first construct a block encoding unitary of $(M^TN)^T M^T N $. For simplicity, let $K \equiv (M^TN)^T M^T N$. It is worth  noting that the entries $K_{ij}$'s are exactly the elements in the $i$-th column of $M^T N$ multiplied with those in the $j$-th column of $M^T N$. Since $M$ and $N$ are not necessarily symmetric, we perform the same trick suggested in~\cite{harrow2009quantum}, that is, instead of working with $M$ and $N$, we work with their Hermitian embedding matrix $\Tilde{M}$ and $\Tilde{N}$ (both of dimension $(n+k) \times (n+k)$) via
\begin{align}
     \Tilde{M}=\begin{pmatrix}
    0 & M \\
    M^T & 0
    \end{pmatrix}, \quad
     \Tilde{N}=\begin{pmatrix}
    0 & N \\
    N^T & 0
    \end{pmatrix}.
\end{align}

It is straightforward to see that
\begin{align*}
    \Tilde{M} \Tilde{N} = \begin{pmatrix}
    MN^T & 0 \\
    0 & M^T N
    \end{pmatrix}. \\
\end{align*}
Without loss of generality, we assume that both $(n+k)$ and $k$ to be a power of 2 (since we can always pad them  with extra zero entries), and let $(n+k) = 2^p k$,  which is due to our assumption that both $(n+k)$ and $k$ are both some powers of 2. Before proceeding further, we note the following: if $\ket{i}$ is $i$-th computational basis state, then $A \ket{i}$ is the $i$-th column of the matrix $A$. Note that lemma \ref{lemma: matrixapp} allows us to `apply' $A$ efficiently, resulting in $A\ket{i}$ entangled with $\ket{0^m}$. Therefore, if $\ket{j}$ is some $j$-th basis state in the Hilbert space of dimension $k$, then $\Tilde{M} \Tilde{N} \ket{2^p}\ket{j} = \ket{2^p} M^TN \ket{j}$. Therefore, by virtue of lemma \ref{lemma: matrixapp}, we have an efficient procedure to implement the following unitary: 
\begin{align}
    U_K \ket{ \bf{0}}\ket{j} = \ket{0^m}\ket{0^m} \ket{2^p} (M^T/s) N\ket{j} + \ket{Garbage},
    \label{eqn: state}
\end{align}
where $s = s_M s_N$, i.e., the product of the sparsity of $M$ times the sparsity of $N$. One may wonder why there are two registers $\ket{0^m}$'s. It comes from the fact that we consecutively use lemma \ref{lemma: matrixapp} to apply $\Tilde{M}$ and $\Tilde{N}$. In order to construct the unitary encoding of $K$, we need to slightly modify $U_K$ to obtain different unitaries $U_{K,1}$ and $U_{K,2}$ in the following ways. \\

$\bullet$ Beginning with $U_k \ket{\bf{0}}\ket{j}$, we append another ancilla initialized in $\ket{00}$:
\begin{align*}
    \ket{0^m}\ket{0^m} \ket{2^p} (M^T N/s) \ket{j} \ket{00}  + \ket{Garbage} \ket{00}.
\end{align*}

Use $X$ gate on the last bit to flip $\ket{0}$ to $\ket{1}$, and  apply $C^{2m+1} X$ conditioned on the first two registers   being $\ket{0^m}\ket{0^m}$ and last bit being $\ket{1}$ (as the control) to flip the next-to-last qubit (as the target) to obtain: 
\begin{align*}
    \ket{0^m}\ket{0^m} \ket{2^p} (M^T N/s) \ket{j} \ket{01}  + \ket{Garbage} \ket{11}.
\end{align*}
We denote the above process as $U_{K,1}$. \\

$\bullet$ With the state  above, we apply a CNOT using the next-to-last bit  as the control bit to flip the last bit, and we obtain:
\begin{align*}
    \ket{0^m}\ket{0^m} \ket{2^p} (M^T N\ket{j}/s) \ket{01}  + \ket{Garbage} \ket{10}.
\end{align*}
We denote the combination of $U_{K,1}$ plus the above extra step as $U_{K,2}$. It is easy to see the following:
\begin{align}
    \bra{ \bf{0}} \bra{i} U_{K,2}^\dagger U_{K,1} \ket{\bra{0}} \ket{j} = \frac{1}{s^2} \bra{i}(M^T N)^T M^T N \ket{j} = K_{ij}/ s^2.
\end{align}

The above property matches with the definition of block encoding~\ref{def: blockencode}. Therefore, we have that $U_{K,2}^\dagger U_{K,1}$ is the unitary block encoding of $K \equiv (M^TN)^T M^T N$ divided by $s^2$.

\subsection{Computing Distance $d_{M_A,M_B}$}
\label{sec: computingdistance}
As we have already succeeded in producing the unitary block encoding of $K$, it is quite straightforward to apply the tools in Lemma~\ref{lemma: qsp}  to simulate $\exp(-i K t)$. We have the following result.
\begin{lemma}
    The evolution $\exp(- i (M^TN)^T M^TN t)$  can be simulated to accuracy $\epsilon_H$ in time 
    \begin{align}
        \mathcal{O}\Big(s^2 log(n+k) (t + log(\frac{1}{\epsilon_H}) )  \Big), 
    \end{align}
\label{lemma: simulation}
where $s = s_M s_N$. 
\end{lemma}

The ability to simulate $\exp(-\textcolor{red}{i} (M^TN)^T M^TN t) $, combined with the QPE subroutine~\cite{kitaev1995quantum, brassard2002quantum} yields the ability to extract the eigenvalues of $(M^TN)^T M^TN $. This method has been used in the famous HHL algorithm~\cite{harrow2009quantum} to invert a given matrix, hence yielding an efficient quantum algorithm for solving linear systems. In order to compute the Grassmann distance, we need to make some modifications, as we will work with mixed states instead of pure states as in~\cite{harrow2009quantum}. However, the analysis regarding error provided in~\cite{harrow2009quantum} still holds in the general case, and we will soon exploit it to prove the efficiency and error bound of our distance calculation procedure. 

The distance calculation procedure begins with the following mixed state:
\begin{align}
    \rho = \frac{1}{k} \sum_{i=1}^k \ket{i}\bra{i}, 
\end{align}
which can be easily prepared by applying $H^{\log (k)} \otimes I^{\log (k)}$ to $\ket{0}^{\log (k)} \otimes \ket{0}^{\log (k)}$, followed by CNOT layers composed of $\log (k)$ CNOT gates, then tracing out either register. Since $\rho$ is diagonal and proportional to the identity matrix in the computational basis state, it is also diagonal in the basis containing eigenvectors of $(M^TN)^T M^TN$. We recall that since $(M^TN)^T M^TN$ is symmetric, its eigenvectors are mutually orthogonal. Let $\{ \ket{u_i}, \lambda_i \}_{i=1}^k $ denotes these eigenvectors and eigenvalues. We use the following important relations and formulas:  \\

$\bullet$ $\lambda_i = \sigma_i^2$ for all i, where $\sigma_i$ is the singular value of $M^TN$.  \\

$\bullet$ $ 0 \leq \sigma_i \leq 1$ for all $i$, i.e., $\sigma_i$ is non-negative for all $i$. \\

$\bullet$ Grassmann distance: 
\begin{align}
    d_{M_A, M_B} = \sqrt{ \sum_{i=1}^k \theta_i^2 },
\end{align}
where $\cos (\theta_i) = \sigma_i$. We remind the readers that  since $ 0 \leq \sigma_i \leq 1$, $\theta_i = \arccos(\sigma_i)$ can be chosen to lie within the range $(0, \pi/2)$ for all $i$.  Therefore, the Grassmann distance can be rewritten as:
\begin{align}
    d_{M_A, M_B} = \sqrt{ \sum_{i=1}^k ( \arccos (\sigma_i)  )^2 } = \sqrt{ \sum_{i=1}^k ( \arccos ( \sqrt{\lambda_i} )  )^2 }.
\end{align}

In order to compute the above distance, we first run the QPE, with varied time unitary $\exp(- i(M^TN)^T M^TN t/C^2) $ and $\rho$ as the input, in a similar fashion to the first part of the HHL algorithm~\cite{harrow2009quantum}. More precisely, we make use of lemma~\ref{lemma: simulation} to apply the controlled evolution (as  in~\cite{harrow2009quantum})
\begin{align}
    \sum_{\tau} \ket{\tau}\bra{\tau} \otimes \exp(- i(M^TN)^T M^TN \tau t_0/ T),
    \label{eqn: t0}
\end{align}
for varying $\tau$ (the register that holds $\ket{\tau}$ is called the phase register) and a fixed $T$ to the following state as part of the QPE subroutine:
\begin{align}
    \frac{1}{T} \sum_{\tau,\tau' = 0}^{T-1} \ket{\tau}\bra{\tau'} \otimes \rho,
\end{align}
followed by an inverse quantum Fourier transform on the phase register. Ideally, if the QPE is exact, we would obtain the following state:
\begin{align}
    \frac{1}{k} \sum_{i=1}^k \ket{\lambda_i}\bra{\lambda_i} \otimes \ket{u_i}\bra{u_i}.
    \label{eqn: phase}
\end{align}

Now we append another ancilla initialized as $\ket{0}$ (technically, it should be written $\ket{0}\bra{0}$ as we are dealing with mixed states; however, it does not possess any systematic issue), performing the following rotation controlled by the phase register $\{ \ket{\lambda_i} \}$
\begin{align}
    \ket{0} \rightarrow \Big( \frac{ \arccos (\sqrt{\lambda_i}) }{ (\pi/2)} \ket{0} + \sqrt{1 -  (\frac{ \arccos (\sqrt{\lambda_i}) }{ (\pi/2)})^2 } \ket{1}  \Big),
    \label{eqn: rotation}
\end{align}
 we obtain:
\begin{align}
    \frac{1}{k} \sum_{i=1}^k \ket{\lambda_i}\bra{\lambda_i} \otimes \ket{u_i}\bra{u_i} \otimes 
    \Big( \frac{ \arccos (\sqrt{\lambda_i}) }{ (\pi/2)} \ket{0} + \sqrt{1 -  (\frac{ \arccos (\sqrt{\lambda_i}) }{ (\pi/2)})^2 } \ket{1}  \Big) \Big( \frac{ \arccos (\sqrt{\lambda_i}) }{ (\pi/2)} \bra{0} + \sqrt{1 -  (\frac{ \arccos (\sqrt{\lambda_i}) }{ (\pi/2)})^2 } \bra{1}  \Big).
\end{align}

The above state seems somewhat complicated. However, we only need to pay attention to the part entangled with the state $\ket{0}\bra{0}$ on the ancilla. If we make a measurement on the ancilla, the probability of measuring $\ket{0}\bra{0}$ is:
\begin{align}
    p_{0} = \frac{4}{\pi^2 k}  \sum_{i=1}^k( \arccos (\sqrt{\lambda_i})  )^2 =  \frac{4}{\pi^2 k} d^2_{M_A, M_B}.
    \label{eqn: p0}
\end{align}

Hence, once we can estimate $p_0$, for example, by repeating the measurement and counting the frequency of seeing 0, then we can estimate $d_{M_A,M_B}$. In order to estimate $p_0$ to accuracy $\delta^2$, and  $d_{M_A,M_B}$ to accuracy $\mathcal{O}(\delta)$, we need to repeat the measurement $\mathcal{O}(1/\delta^2)$ times. 

\subsection{Error Analysis}
The previous section assumes an ideal case, as we have emphasized, that when the QPE is exact. In reality, there is an error due to finite-bit precision in the QPE, which means that we only obtain the approximated phase, $\Tilde{\lambda_i}$, instead of $\lambda_i$. Fortunately, this critical issue has been analytically worked out and dealt with in~\cite{harrow2009quantum}. We strongly refer the readers to~\cite{harrow2009quantum} for a complete treatment, as we will not repeat the calculation here. Rather, we directly apply their analysis to our context. Denote $\kappa$ as the conditional number (basically the ratio of maximum singular value over minimum singular value, in magnitude)  of $M^TN $ (which means that $\kappa^2$ is the conditional number of $(M^T N)^T M^T N$), if we have  
$$ t_{0} = \mathcal{O} (\kappa^2/\epsilon_p), $$
where $t_0$ is a fixed term appearing in the conditional evolution, i.e., in Eqn.~\ref{eqn: t0}. Then the analysis from~\cite{harrow2009quantum} applied here yields the following bound on the error:
\begin{align}
\sqrt{ \frac{4}{\pi^2 k} \sum_{i=1}^k \Big(  \arccos (\sqrt{\Tilde{\lambda_i}})  -  \arccos( \sqrt{\lambda_i} ) \Big)^2 }  \leq \epsilon_p.
\end{align}

The finite-bit precision in the QPE further affects the actual measurement outcome. We summarize our result in the following lemma.
\begin{lemma}
    In non-ideal case, the probability of measuring $\ket{0}\bra{0}$ now becomes:
\begin{align}
    \Tilde{p}_0 =  \frac{4}{\pi^2 k}  \sum_{i=1}^k( \arccos (\sqrt{ \Tilde{\lambda_i}})  )^2,
    \label{eqn: nonidealp0}
\end{align}
and 
\begin{align}
    \Big| \Tilde{p}_0 - p_0  \Big| \leq  \epsilon_p. 
\end{align}
\label{lemma: errorbound}
\end{lemma}
Proof of the above lemma is given in Appendix~\ref{sec: errordistance}. It essentially means that in the non-ideal case, we can only estimate an approximated value of $p_0$, or equivalently, $d_{M_A,M_B}$, as $d_{M_A,M_B} = \frac{1}{2} \pi \sqrt{k} \sqrt{p_0} $.  \\

As we have outlined a quantum algorithm for computing linear subspace distance, we have seen many sources of error throughout the whole procedure. For the purpose of summary, we revise the main steps of our algorithm with the corresponding error contribution from such steps before establishing the final running time with respect to overall error tolerance, for which we eventually set to $\epsilon$: \\

$\bullet$ \textbf{\textit{State Preparation:}} Error $\epsilon_S$ induced from matrix application step~\ref{lemma: matrixapp}. In principle, $\epsilon_S$ would contributes to the simulation of error of $\exp(-\textcolor{red}{i} (M^TN)^T M^TN t)$. However, the running time of the preparation step scales as $\mathcal{O}(\log (1/\epsilon_S))$, which is very efficient. Therefore, we can neglect the error contribution from this step, as we can make it very small at a modest cost. \\

$\bullet$ \textbf{\textit{Simulating Evolution:}} Error $\epsilon_H$ induced directly from improper simulation of $\exp(- \textcolor{red}{i}(M^TN)^T M^TN t)$, as a result of density matrix exponentiation technique~\cite{lloyd2014quantum}. As we stated in lemma~\ref{lemma: simulation}, the running time scales $\mathcal{O} (log (1/\epsilon_H))$, which is logarithmic and hence, is efficient. We remind that our algorithm shares a similar routine as the HHL algorithm~\cite{harrow2009quantum}, but in their work, the error induced from simulating the evolution of a given matrix, say, $\exp(-iHt)$ is negligible. \\

$\bullet$ \textbf{\textit{Performing Quantum Phase Estimation:}} The error  is resulted from the improper phase estimation $\epsilon_P$. This is probably the most complicated factor in our algorithm. Inaccurate phase estimation results in an inaccurate distance formula, as we point out in equation~\ref{eqn: nonidealp0}. \\

$\bullet$ \textbf{\textit{Estimating Distance $d_{M_A,M_B}$ From Measurement Outcome:}} Its error $\delta$ comes from the estimation of $\Tilde{p_0}$, which gives a further running time $\mathcal{O}(1/\delta^2)$ as a result of the Chernoff bound. We remark that the quantum amplitude estimation~\cite{brassard2002quantum} can improve the cost to $\mathcal{O}(1/ \delta)$. This step can be done with $\mathcal{O}(1)$ time by preparing $\mathcal{O}(1/\delta^2)$ identical quantum circuits and running them in parallel before averaging the result statistically.  \\ 

All in all, the most dominant error comes from the phase estimation step, similar to~\cite{harrow2009quantum}. For simplicity, we can set the desired error tolerance to be $\epsilon$. We establish our first main result.
\begin{theorem}[Estimation of Grassmann Distance]
\label{thm: grassmandistance}
Given access to matrix $M$ and $N$ $\in R^{n \times k}$ as defined in section~\ref{sec: classical}. Let $\kappa$ denote the conditional number of $M^T N$. Then the Grassmann distance between $M_A$ and $M_B$, which is spanned by column vectors of $M$ and $N$, respectively, can be estimated to additive accuracy $\epsilon$ in time
\begin{align*}
    \mathcal{O}\Big( \kappa^2 s^2 \log (n+k) \cdot \frac{1}{\epsilon^2} \Big).  
\end{align*}
\end{theorem}

This is a general statement about the running time of the algorithm that we have outlined. Now, we will discuss some aspects of the algorithm, and particularly, we will show that there are specific scenarios that achieve a better running time.

\subsection{Comments}
\noindent\textit{\textbf{Conditional Number:}} The dependence on conditional number $\kappa$ is worth examining further. The concern is, when will $\kappa$ be small? Since $M$ and $N$ are formed by orthogonal vectors that could be drawn from random, it is generally hard to predict how $\kappa$ would behave. However, in the above paragraph, we have described a geometric picture about the entries of $M^T N$, which is essentially the inner product of corresponding columns of $M$ and $N$. We now mention the following interesting result regarding the conditional number of a matrix. 

\begin{theorem}[Theorem 1 in \cite{varah1975lower} (Lower Bound on Singular Value)]
    Let $A$ be $n\times n$ matrix and assumed to be diagonally dominant by rows and set $\alpha = \min_k ( |a_{kk}| - \sum_{j\neq k} |a_{jk}|)$. Then $\sigma_{min} > \alpha$. 
\end{theorem}

The above result can equivalently work if $A$ is diagonally dominant by columns. Recall that the diagonal entries $(M^TN)_{ii}$ are the inner product of $i$-th column of $M$ and $N$. What does it mean for $M^TN$ to be diagonally row/column dominant? If for every $i$, we have the condition that $m_i$ is `very close' to $n_i$, and quickly becomes almost orthogonal to the remaining vector from $\{n\}$ as the index runs away from $i$, then we can guarantee that the matrix $M^T N$ is diagonally row dominant. By virtue of the above theorem, we have that the minimum singular value of $M^T N$ is lower bounded by a constant (independent of dimension), which directly implies that its conditional number $\kappa$ is upper bounded by a constant.  \\

\noindent\textit{\textbf{Advantage Over Classical Algorithm:}} We have mentioned in Section \ref{sec: classical} that the seemingly best classical algorithm for computing Grassmann distance takes time 
\begin{align}
    \mathcal{O}(n^2 k + k^3 ).
\end{align}
Compared with the quantum running time~(Theorem~\ref{thm: grassmandistance}), we see that if $\kappa, s \ll n,k$, for example, are of order $\approx \mathcal{O}( \log(n,k))$, then there is an exponential speedup with respect to both $n$ and $k$. In the previous paragraph, we have mentioned a setting where $\kappa$ can be small. Recall that $s = s_M s_N$, and, therefore, our quantum algorithm performs the best when both $M$ and $N$ are sparse. In the denser regime, i.e., $s \approx \max(n,k)^2$, and this means that with respect to $(n,k)$, the quantum running time could be as much as $\mathcal{O}(\max(n,k)^4)$, which is larger than the classical algorithm. If $s$ grows with fractional power w.r.t $\max(n,k)$, e.g, $\max(n,k)$, $\sqrt{\max(n,k)}$ or $\max(n,k)^{1/3}$, then we would achieve corresponding quadratic or polynomial speedup. The above running time holds for all cases. Because even when both $M$ and $N$ are sparse, their product is not guaranteed to be sparse. Therefore, one may still need to perform SVD on a dense matrix.

\section{Quantum Algorithm for Computing Ellipsoid Distance}
\label{sec: quantumellip}

Now we turn our attention to the second problem: computing the distance between two ellipsoids $\mathcal{E}_M$ and $\mathcal{E}_N$, each defined by a real symmetric positive definite matrix $M$ and $N$ $\in R^{n \times n}$, respectively. Without loss of generality, we assume eigenvalues of $M$ and $N$ to be bounded in the known range ($1/\kappa, 1$), which is always achievable by rescaling. To recall, such distance is calculated as follows:
\begin{align}
    \delta_{M,N} = \sqrt{ \sum_{i=1}^n \log^2 (\lambda_i(M^{-1}N)) }.
\end{align}
While at first, the symmetric property of both $M$ (and hence of $M^{-1}$) and $N$ may seem useful; however, this does not make calculating the above distance trivial. The reason is that $M^{-1}N$ is not necessarily symmetric, which can be seen by simply performing the transpose:
\begin{align*}
    (M^{-1}N)^T = N^T (M^{-1})^T = N M^{-1},
\end{align*}
which is different from $M^{-1}N$ in general. 

\subsection{Encoding Matrix}

For convenience, we first set $P = M^{-1}N$. In order to resolve the non-symmetric issue, we use the same trick as we did in last section, i.e., we would try to simulate $\exp(-i P^T P t )$, then doing algebraic operations on its eigenvalues. 
This problem is somewhat more challenging than the Grassmann distance, as there is a requirement for matrix inversion. This is exactly the utility of the celebrated quantum linear solver algorithm~\cite{harrow2009quantum, childs2017quantum}. We remark that, in our case, $M$ can be dense. The inversion of a dense matrix has been done in~\cite{wossnig2018quantum}, with a particular quantum data structure. That kind of quantum data structure is not assumed in our case, as we only work with the familiar blackbox model. Therefore, we would use the result of~\cite{childs2017quantum} to achieve the matrix inversion since this method achieves better scaling time on error tolerance. 

\begin{lemma}[Matrix Inversion \cite{childs2017quantum}]
\label{lemma: matrixinversion}
Given access to some Hermitian matrix M of size $n \times n$, an initial state $b \equiv \ket{b}$. There is a unitary that performs the following map:
\begin{align}
    U_M \ket{0}\ket{b} = \ket{0} |M^{-1} b| \ket{M^{-1}b} + \ket{1}\ket{Garbage}.
\end{align}

The running time of $U_M$ is $\mathcal{O}\Big( s_M \log (n) \kappa_M  poly(\log 1/\epsilon) )   \Big)$ where $s_M$ is the sparsity of matrix M, $\epsilon$ is some tolerance error (the state $|M^{-1} b| \ket{M^{-1}b}$ is just an approximation of the state). 
\end{lemma}

With the above result and combined with lemma~\ref{lemma: matrixapp}, we are able to create the state that corresponds to columns of $P$. More specifically, we use Lemma~\ref{lemma: matrixapp} to achieve the following:
\begin{align}
    U_N \ket{0^m} \ket{i} = \ket{0^m} (N/s_N)  \ket{i} + \ket{Garbage},
    \label{eqn: eq25}
\end{align}
where $s_N$ is the sparsity of $N$. For a reason that would be clear later, we append another ancilla $\ket{0}$ and work in a larger Hilbert space. We would have instead: 
\begin{align}
  \mathbb{I} \otimes  U_N \ket{0}\ket{0^m} \ket{i} = \ket{0} \ket{0^m} (N/s_N)  \ket{i} + \ket{0}\ket{Garbage}.
\end{align}

Next, we use $\ket{0^m}$ as the control system to flip the first qubit, i.e., we transform the above state to: 
$$ \ket{1} \ket{0^m} (N/s_N)  \ket{i} + \ket{0}\ket{Garbage}.$$
We then use matrix inversion from Lemma~\ref{lemma: matrixinversion}, controlled by the first $\ket{1}$  to invert $M$ on the state $(N/s_N) \ket{i}$. To be more precise, we need $\ket{0}(N/s_N)  \ket{i}$ in order for $U_M$ to be effective. The extra $\ket{0}$ can be borrowed from $\ket{0}^m$, i.e, we have $\ket{0}^m (N/s_N)  \ket{i} \equiv \ket{0}^{m-1} \ket{0}(N/s_N)  \ket{i}$. We further note that this time the unitary $U_M$ in \ref{lemma: matrixinversion} is controlled by a qubit being $\ket{1}$ (the first register in the above state).  We obtain the following unitary denoted as $U$:
\begin{align}
    U \ket{0}\ket{0^m} \ket{i} = \ket{1} \ket{0}^{m-1}\Big( \ket{0} M^{-1} (N/s_N) \ket{i} + \ket{1}\ket{Garbage_1} \Big) + \ket{0}\ket{Garbage}.
\end{align}

Since the role of all garbage states is the same, as they do not contribute to the final result, we simplify the above representation as: 
\begin{align}
 \ket{1} \ket{0}^m M^{-1} N/s_N \ket{i} + \ket{Garbage}.
\end{align}
Now, we again use the $m$-qubit register as control and, conditioned on they being $\ket{0^m}$, flip the first qubit back to $\ket{0}$.  The final state becomes
\begin{align}
    \ket{0} \ket{0^m} M^{-1} (N/s_N) \ket{i} + \ket{Garbage}, 
\end{align}
which has a similar form as~\ref{eqn: state}. Therefore, we can use the same procedure, with extra two ancilla qubits, as we outlined in section~\ref{sec: preparingstate} to block encode matrix $C^2 P^T P /s_N^2$ (recall that $M^{-1}N\ket{i}$ is the $i$-th column of $M^{-1}N$ and we have set $P \equiv M^{-1}N$). In a clear manner, the application of Lemma~\ref{lemma: qsp} yields the following:

\begin{lemma}
    The simulation of $\exp(-i (M^{-1}N )^T M^{-1}N t)$ can be achieved up to accuracy $\epsilon$ in 
    \begin{align*}
        \mathcal{O} \Big( s_M s_N^2 poly\log (n , \frac{1}{\epsilon}) \cdot \kappa_M t  \Big).
    \end{align*}
    \label{lemma: lemma8}
\end{lemma}

\subsection{Computing Ellipsoid Distance}
\label{sec: comped}

Once we can simulate $\exp(-i P^T P t)$, we then run the QPE with completely mixed state $\rho = (1/n) \sum_{i=1}^n \ket{i}\bra{i}$ as the input. After the QPE, we would obtain a state similar to Eqn.~\ref{eqn: phase}. We then append an ancilla $\ket{0}$ and rotate as the following (again, we are ignoring the mixed state formalism):
\begin{align}
    \ket{0} \rightarrow  \log( \Tilde{\lambda_i} ) \ket{0} + \sqrt{1 - \log ( \Tilde{\lambda_i})^2} \ket{1},
\end{align}
where $\Tilde{\lambda_i}$ is the approximated $i$-th eigenvalue of $P \equiv M^{-1}N$. We then measure the ancilla, with the probability of measuring $\ket{0}$ could be shown to be:
\begin{align}
    p = \sum_{i=1}^n \log^2 (\Tilde{\lambda_i})/ n  = \Tilde{\delta}_{M,N}^2 / n,
\end{align}
where $\Tilde{ \delta}_{M,N}^2$ denotes the approximated value of the real ellipsoids distance. Following the same analysis as in previous section (which is based on analysis in the HHL algorithm), if we choose $t = \mathcal{O}(\kappa_P/\epsilon)$ (where $\kappa_P$ is the conditional number of P) in the simulation of $\exp(-iP^T Pt)$, then we can guarantee an overall error $\epsilon$, i.e, $|  \Tilde{ \delta}_{M,N} - \delta_{M,N}| \leq \epsilon$. The value of $p$ itself can be estimated to accuracy $\epsilon$ by repeating the measurement $\mathcal{O}(1/\epsilon^2)$ times, which can be improved to $\mathcal{O}(1/ \epsilon)$ by employing quantum amplitude estimation~\cite{brassard2002quantum}. We now establish our results regarding the estimation of the ellipsoid distance.

\begin{theorem}[Estimation of Ellipsoids Distance]
Let $\mathcal{E}_M$ and $\mathcal{E}_N$ be ellipsoids defined by two real symmetric positive definite matrix M and N. Given local access to entries of M and N, the distance between $\mathcal{E}_M$ and $\mathcal{E}_N$, denoted as $\delta_{M,N}$, can be estimated in time:
\begin{align*}
        \mathcal{O} \Big( s_M s_N^2 \log (n) \frac{\kappa_P \kappa_M}{\epsilon^2}  \Big), 
\end{align*}  
where $\kappa_P$ is the conditional number of $M^{-1}N$. 
\end{theorem}

 One may wonder that there seems to be a missing factor of order $\mathcal{O}(polylog(1/\epsilon)$, as it appears in the running time of Lemma \ref{lemma: lemma8}. The main reason is that we simply absorb that scaling into the $\mathcal{O}(1/\epsilon^2)$, which is a result of Hadamard estimation plus an HHL-like procedure. \\

\textit{\textbf{Advantage Over Classical Algorithm:}} As we have mentioned from Sec.~\ref{sec: classical}, the best classical algorithm for computing ellipsoids distance has running time $\mathcal{O}(n^3)$, with the running time dominated by performing classical SVD. If both $M$ and $N$ are sparse, with low conditional numbers (of order $\log (n) $), then there is exponential speedup. Even when matrix $M$ is not sparse, i.e., $s_M \in \mathcal{O}(n)$, but $N$ is sparse and $M^{-1}N$ has low conditional number, there is polynomial speedup. \\ 

In the Grassmann distance case, we have pointed out the specific scenario where the conditional number of the corresponding matrix could be low; however, it is quite difficult to find such a case in the ellipsoid distance case. The reason is apparently due to the inversion $M^{-1}$. If $M$ is orthogonal, i.e., $M^{-1} = M^T$, then it becomes similar to the  Grassmann distance case, where we can have a specific scenario with a low conditional number. 

\section{Extension to Different Kind of Distances}
\label{sec: extension}

The above two main quantum algorithms were devoted to estimating the so-called Grassmann distance and ellipsoid distance. Here we aim to extend the application of our quantum algorithm, specifically the Grassmannian case, by discussing some other related distances that could also be practically useful~\cite{ye2016schubert}. We note that we would adopt the same notations as in section~\ref{sec: quantum}. \\

The first one is called \textit{Asimov distance}, which is defined as following: 
\begin{align}
    d^A_{M,N} = \theta_k,
\end{align}
where $\cos (\theta_k) = \sigma_k$ is the smallest singular value of $M^T N$. Therefore, we first need to find the value of $\sigma_k$. Finding the maximum and minimum eigenvalues of a given Hermitian matrix has been done in~\cite{nghiem2022quantum}, based on the (classical) power method. In our case,  we are given the ability to simulate $\exp( -i (M^T N)^T M^T N t )$. Still, the method of~\cite{nghiem2022quantum} can be adapted in a straightforward manner. It will yield us the approximated minimum eigenvalue of $(M^T N)^T M^T N$, which in turn gives an estimation of $\sigma_k^2$. As was analyzed in~\cite{nghiem2022quantum}, the random initialization step is quite critical in the performance of the algorithm, as it might diminish the exponential speedup. In general, only quadratic speedup is obtained.   \\

The next one, very close to  the Asimov distance, is called \textit{projection distance}, which is defined simply as:
\begin{align}
    d^P_{M,N} = \sin (\theta_k), 
\end{align}
where the angle $\theta_k$ is defined in the same way as in the Asimov distance case. 
It is easy to see that $(d^P_{M,N})^2  = 1 - \cos (\theta_k)^2 = 1- \sigma_k^2$. Therefore, estimating $\sigma_k^2$ is sufficient. Such estimation of $\sigma_k$ was just described above by using the method in~\cite{nghiem2022quantum}. Thus, its efficiency is the same as that of the Asimov distance.\\

Now we discuss the \textit{Chordal distance}, which is defined as follows:
\begin{align}
    d^C_{M,N} = \Big( \sum_{i=1}^k \sin (\theta_i)^2 \Big)^{1/2}.
\end{align}

We note the following: 
\begin{align}
    (d^C_{M,N})^2 =&  \sum_{i=1}^k \sin (\theta_i)^2   \\
            & =  k - \sum_{i=1}^k \cos (\theta_i)^2  \\
            & = k - \sum_{i=1}^k \sigma_i^2.
\end{align}
Therefore, it suffices to estimate $\sum_{i=1}^k \sigma_i^2$. This can be done in a very similar manner to Grassmann distance estimation, except that in the rotation step \ref{eqn: rotation}, we do not have to do any further arithmetic operation, which makes it even simpler. Therefore, the estimation of the Chordal distance can be done with the same running time as that of the Grassmann distance; see Theorem~\ref{thm: grassmandistance}. \\

\section{Estimating Grassmann Distance and Ellipsoid Distance in The Memory Model }
\label{sec: memorygrassman}
In the above sections, we work in the standard blackbox model that assumes coherent access to entries of corresponding matrices. Here, we explore how the so-called memory model can potentially enhance the estimation of Grassmann and ellipsoid distances. Particularly, we shall see that the quantum running time in the memory model is sparsity-independent. This is within our expectation, as in~\cite{wossnig2018quantum}, the authors also used such a model to construct a quantum linear solver that has running time independent of the sparsity of the given matrix.  \\

To begin with, the memory model was proposed in~\cite{kerenidis2016quantum} as a novel quantum architecture that allows efficient load out of classical data. In the standard blackbox model, data entries are accessible individually, whereas, in the memory model, data is usually loaded column/row-wise. In~\cite{kerenidis2016quantum}, the authors showed how this model can give rise to an efficient quantum algorithm for the recommendation system. In particular, this model, combined with the famous quantum phase estimation algorithm, yields a sparsity-independent quantum linear solver~\cite{wossnig2018quantum}. While the problem of recommendation system was ``de-quantized" in the seminal work of Tang \cite{tang2019quantum} (where the author showed that under appropriate assumption regarding input access, there exists a classical algorithm that could solve the recommendation system problem with at most a polynomial slowdown compared to the quantum counterpart), the advantage of a dense quantum linear solver seems to hold due to BQP-completeness of matrix inversion~\cite{harrow2009quantum}. Throughout this work, we have seen that the blackbox model could yield an efficient quantum algorithm for estimating different kinds of geometric distances. Thereby, it is very interesting to expand the potential of the memory model in solving various computational tasks other than dense linear systems~\cite{wossnig2018quantum}. \\

Before diving into the algorithms, we first recall the features of the memory model. 
\begin{lemma}[Data Structure \cite{kerenidis2016quantum, wossnig2018quantum}]
\label{lemma: datastructure}
    Given a matrix $M \in \mathbb{R}^{n \times m}$. Then there exists a quantum data structure that allows the following coherent mapping in $\mathcal{O}(poly(\log(mn)))$ time: 
    \begin{align*}
        U_M \ket{\bf i}\ket{0} \rightarrow \frac{1}{||A_i||} \sum_{j=1}^n A_{ij}\ket{j}\ket{i}, \\
        U_N \ket{0}\ket{j} \rightarrow \frac{1}{||A||_F} \sum_{i=1}^m ||A_i|| \ket{i} \ket{j},
    \end{align*} 
    where $i,j$ refers to the row and column index of A. 
\end{lemma}

Similar to our previously described algorithm in the blackbox model, we will employ the above mapping to construct the block encoding of corresponding matrices, followed by the QPE and measurement to extract the desired distances. 

\subsection{Grassmann Distance}
Recall that in the Grassmann distance problem, the algorithm outlined in section~\ref{sec: quantum}  relies on the simulation of $\exp( -i (M^TN)^T M^T N t)$. The question now is how to perform block-encoding of $M^TN$ given the memory model structure in Lemma~\ref{lemma: datastructure}. We first remind that the matrix $M^T $ is of size $k \times n$, and the matrix $N$ is of size $n \times k$, where $n$ is the dimension of given data and $k$ is the number of data points. WOLG, for simplicity, we embed $M$ and $N$ into a bigger matrix of size $\max (k,n) \times \max (k,n)$, with those additional entries set to 0. With such simplification, we would work with a square matrix instead, which is more convenient. \\

Now, we attempt to do block-encoding of $M^T N$ given the memory model structure. We note that, as mentioned in \cite{wossnig2018quantum}, the memory model naturally provides the block-encoding of $M/|M|_F$ and $N/|N|_F$ (we shall derive this in the appendix), where $| . |_F$ refers to the Frobenius norm of these matrices. For convenience, we denote those block-encoding unitaris as $U_M$ and $U_N$, respectively. \\

We first remind from definition \ref{def: blockencode} that if a matrix $A$ is encoded in some unitary $U$, then it can be written as: 

\begin{align}
    U = \ket{ \bf{0}}\bra{ \bf{0}} \otimes A + \cdots
\end{align}
It is worthy to observe that, the above unitary $U$ acts on the state $\ket{\bf 0}\ket{\phi}$, where $\ket{\phi}$ has the same dimension as $A$, as following: 
\begin{align}
    U \ket{\bf 0} \ket{\phi} = \ket{\bf 0} A\ket{\phi} + \sum_{j \neq \textbf{0}} \ket{j} \ket{\phi_j},
\end{align}
where $\ket{\phi_j}$ refers to some redundant state. For some reason that would become clear later, we add an ancilla $\ket{0}$ and produce the following:
\begin{align}
    \mathbb{I} \otimes U \ket{0} \ket{\bf 0} \ket{\phi} =\ket{0} \ket{\bf 0} A\ket{\phi} + \ket{0} \sum_{j \neq \textbf{0}} \ket{j} \ket{\phi_j}.
\end{align}

With the above observation, let $\ket{i}, \ket{k}$ be some arbitrary computational basis states (with their dimensions corresponding to those of the matrices $M,N$, respectively), we have the following:
\begin{align}
    \mathbb{I} \otimes U_M \ket{0} \ket{\bf 0} \ket{i} = \ket{0} \ket{\bf 0} (M/|M|_F) \ket{i} + \ket{0} \sum_{j \neq \textbf{0}} \ket{j} \ket{\phi_j} \equiv \ket{\Phi_M}, \\
    \mathbb{I} \otimes U_N \ket{0} \ket{\bf 0} \ket{k} = \ket{0} \ket{\bf 0} (N/|N|_F) \ket{k} + \ket{0} \sum_{j \neq \textbf{0}} \ket{j} \ket{\phi_j} \equiv \ket{\Phi_N}.
\end{align}

$\bullet$ Now for state $\ket{\Phi_N}$, we use $X$ gate to flip the ancilla, e.g., the first qubit to obtain 
\begin{align}
   X \otimes \mathbb{I} \ket{\Phi_N} = \ket{1} \ket{\bf 0} (N/|N|_F) \ket{k} + \ket{1} \sum_{j \neq \textbf{0}} \ket{j} \ket{\phi_j}  \equiv \ket{\Phi^1_N}.
\end{align}

$\bullet$ Now we use the register $\ket{\bf 0}$ as a control register to flip the first qubit back to $\ket{0}$  (i.e., flip conditioned the register being ${\bf 0}$), and we obtain the state:
\begin{align}
    \ket{0} \ket{\bf 0} (N/|N|_F) \ket{k} + \ket{1} \sum_{j \neq \textbf{0}} \ket{j} \ket{\phi_j} \equiv \ket{\Phi^2_N}.
\end{align}

Denote the unitary $\mathbb{I} \otimes U_N$ plus the above two additional steps as $P_N$. As previously pointed out, for arbitrary matrix $A$, $A\ket{i}$ is the $i$-th column of $A$. Furthermore, the entries of matrix $M^T N$ are basically the inner product of columns of $N$ and $M$. It is then straightforward to observe that: 
\begin{align}
    \braket{\Phi_M | \Phi^2_N} = \bra{i} \frac{M^\dagger}{|M|_F} \frac{N}{|N|_F}\ket{k} =  \frac{(M^T N)_{ik}}{|M|_F |N|_F  },
\end{align}
where we use that $M,N$ are real then $M^\dagger = M^T$, which is just the transpose. The above property matches perfectly with the definition of block-encoding~\ref{def: blockencode}. Therefore, the unitary $(\mathbb{I} \otimes U_M^\dagger) (P_N )$ is exactly the unitary block-encoding of $(M^T N)/|M|_F |N|_F$. Note that this unitary has a larger dimension than the initial unitary encoding of $M$ and $N$.\\

Given the block-encoding of $M^T N/ |M|_F |N|_F$, it is very easy to perform the block encoding of $(M^T N)^T (M^T N)/ (|M|_F |N|_F)^2 $. Basically, we can apply the same procedure as we just outlined. Therefore, we do not repeat it here. We simply would proceed with the following lemma:
\begin{lemma}
    In presence of the memory model with given data structure \ref{lemma: datastructure} of $M$ and $N$, the simulation of $\exp( -i(M^TN)^T M^T N t)$ can be achieved up to accuracy $\epsilon$ in time 
    \begin{align*}
        \mathcal{O}\Big(  polylog((n,k)) |N|_F^2 |M|_F^2 (t + \log(1/\epsilon)  )  \Big),
    \end{align*}
\end{lemma}
where we remark that $(n,k)$ refers to $\max(n,k)$. Using such ability, one can apply essentially the same algorithm outlined in \ref{sec: computingdistance} to estimate the Grassmann distance. Therefore, we have the following straightforward statement:

\begin{theorem}
    Given access to matrices $M$ and $N$ in the memory model, the Grassmann distance between $M_A$ and $M_B$ can be estimated to additive accuracy $\delta$ in time
    \begin{align*}
        \mathcal{O}\Big( \kappa^2 |M|_F^2 |N|_F^2 polylog((n,k)) \cdot \frac{1}{\delta^2}  \Big). 
    \end{align*}
\end{theorem}

Now, we make the following comparison. Since both $M$ and $N$ are of size $n \times k$ and are assumed to have columns of unit norm, their Frobenius norm is $\sqrt{k}$. Therefore, the actual quantum running time is $\mathcal{O}\Big( \kappa^2 k^2 polylog((n,k)) \cdot \frac{1}{\delta^2}  \Big) $. If the conditional number $\kappa$ does not grow anything faster than $polylog((n,k))$, this running time is considerably much shorter than the classical running time $\mathcal{O}(n^2 k + k^3)$, as well as the quantum algorithm in the blackbox model (see Sec \ref{sec: computingdistance}) in the dense setting, which could be $\mathcal{O}(k^4)$). 

\subsection{Ellipsoid Distance}
Now, we discuss how ellipsoid distance could be estimated in the memory model. Recall that in the ellipsoids distance problem, we are given two symmetric positive definite matrices $M$ and $N$, and we need to estimate the following quantity: 
\begin{align}
    \delta_{M,N} = \sqrt{ \sum_{i=1}^n \log^2 (\lambda_i(M^{-1}N)) },
\end{align}
For completeness, we first recall a few key operations as well as important observations.  \\

The first one is, as we also pointed out in the last section, the memory model naturally produces the block encoding unitaries of $M/|M|_F$ and $N/|N|_F$, respectively. We denote them here as $U_M$ and $U_N$ for simplicity.    \\

The second one is the operation of $U_M$ (respectively $U_N$) on the given arbitrary state $\ket{\bf 0} \ket{\phi}$: 
\begin{align}
    U \ket{\bf 0} \ket{\phi} = \ket{\bf 0} A\ket{\phi} + \sum_{j \neq \textbf{0}} \ket{j} \ket{\phi_j}.
\end{align}

The last one is the matrix inversion quantum algorithm that was proposed in \cite{wossnig2018quantum}. 

\begin{lemma}
\label{lemma: invertmemory}
    In the presence of the memory model (see \ref{lemma: datastructure}), given some initial state $\ket{b}$, the following unitary could be implemented:
    \begin{align}
        U^M_{invert}\ket{0}\ket{b} = \ket{0} C M^{-1}\ket{b} + \ket{1}\ket{Garbage}.
    \end{align}
    The running time of $U^M_{invert}$ is 
    \begin{align*}
        \mathcal{O}\Big( \kappa_M |M|_F \cdot  polylog(n) \frac{1}{\epsilon}  \Big),
    \end{align*}
    where $\kappa_M$ is the conditional number of $M$; $|M|_F$ is the Frobenius norm of $M$ and $\epsilon$ is the error tolerance; and $C$ is the factor that is required for the normalization purpose, e.g., $C \leq 1/\kappa$. 
\end{lemma}

Now, we are ready to describe our quantum algorithm in the memory model. It turns out that the quantum algorithm in this case (memory model) is essentially similar to that of blackbox model (see Section \ref{sec: quantumellip}). As we shall see, the only difference is the matrix inversion step, where the memory model supports a faster subroutine. \\

We begin with some computational basis state $\ket{i}$ (which shares the same dimension as $M,N$), we use $U_N$ to act and obtain:
\begin{align}
   \mathbb{I} \otimes U_N \ket{0} \ket{\bf 0}\ket{j} = \ket{0} \ket{\bf 0} (N/|N|_F) \ket{i} + \ket{0} \sum_{j \neq \textbf{0}} \ket{j} \ket{\phi_j}.
\end{align}
We make an observation that this state is very similar to Eqn.~\ref{eqn: eq25}.  Following the same procedure as in \ref{sec: quantumellip} (basically the whole paragraph after equation \ref{eqn: eq25}), we use $\ket{\bf 0}$ to flip the first qubit $\ket{0}$ to $\ket{1}$, then use $\ket{1}$ as a controlling qubit to apply Lemma~\ref{lemma: invertmemory}. We yield the following state:
\begin{align}
    \ket{0}\ket{0^m} M^{-1}(N/|N|_F) \ket{i} + \ket{Garbage}.
\end{align}
Therefore, we have the following:
\begin{lemma}
   In the presence of memory model, the simulation of $\exp(-i P^T P t)$ can be achieved up to accuracy $\epsilon$ in
   \begin{align*}
       \mathcal{O} \Big( \kappa_M |M|_F \cdot polylog(n) \frac{1}{\epsilon} t |N|_F^2   \Big). 
   \end{align*}
\end{lemma}
With this, following the same procedure outlined \ref{sec: quantumellip}, the ellipsoid distance can be estimated.

\begin{theorem}[Ellipsoid Distance in Memory Model]
    In the presence of the memory model, the distance between two ellipsoids defined by two real symmetric positive definite matrices $M$ and $N$ can be estimated up to an additive accuracy $\epsilon$ in time:
    \begin{align*}
          \mathcal{O} \Big( |M|_F \cdot polylog(n) |N|_F^2 \frac{\kappa_P \kappa_M }{\epsilon^3}  \Big), 
    \end{align*}
    where $P \equiv M^{-1} N$.
\end{theorem}

\section{Conclusion}
\label{sec:conclusion}
Motivated by fast-pace development of quantum algorithm, as well as the prospect of quantum advantage, we have outlined two quantum algorithms for estimating Grassmann distance and ellipsoid distance between two subspaces formed by corresponding data elements. We have specifically constructed quantum algorithm in both data models, which is the standard blackbox model and the newly developed memory model. Our algorithm is built upon density matrix exponentiation \cite{lloyd2013quantum}, fast quantum matrix application \cite{nghiem2022quantum} and quantum linear solving algorithm \cite{harrow2009quantum, childs2017quantum}. Under the corresponding assumption regarding the input access, as well as appropriate conditions regarding sparsity and conditional number, our quantum algorithms yield significant speedup compared to classical algorithms that could solve the same problems, e.g, estimating Grassmann distance and ellipsoids distance. The novelty of our approach, or more specifically, the use of quantum phase estimation, lies in the way that we perform algebraic operations directly on the phase registers, and essentially combine them to estimate corresponding distances. As the effort for expanding the applicability of quantum computer still goes on, we strongly believe that the techniques outlined here could find more impactful benefits in devising novel quantum algorithm to solve challenging computational problems. As we mentioned, our work have added to a few existing works, e.g, topological data analysis \cite{lloyd2016quantum}, homology detection \cite{nghiem2022constant}, that explores the potential of quantum computing methods in (computational) geometry $\&$ topology, which is a very fundamental, yet having increasingly significant impacts to real-world problems. What else can stem from this work is a completely interesting avenue.  \\

\textbf{Acknowledgement:} The author thanks Tzu-Chieh Wei for careful reading and thoughtful comments on the work. The author also thanks Phenikaa Institute for Advanced Study (PIAS), Phenikaa University for hospitality, where part of the work was done.   This work was supported in part by the US Department of Energy, Office of Science, National Quantum Information Science Research Centers, Co-design Center for Quantum Advantage
(C2QA) under contract number DE-SC0012704. 
We also acknowledge the support from a Seed Grant from
Stony Brook University’s Office of the Vice President for Research.

\bibliography{ref.bib}{}
\bibliographystyle{unsrt}

\clearpage
\newpage
\onecolumngrid

\appendix

\section{Proof of Lemma 4}
\label{sec: errordistance}
Here we provide the proof of Lemma \ref{lemma: errorbound}. To remind, we have:
\begin{align}
\Tilde{p}_0 = \frac{4}{\pi^2 k}\sum_{i=1}^k (\arccos \sqrt{ \Tilde{\lambda}_i})^2, \\ 
p_0 = \frac{4}{\pi^2 k}\sum_{i=1}^k (\arccos \sqrt{ \lambda}_i)^2.
\end{align}

We also have the following:
\begin{align}
\sqrt{ \frac{4}{\pi^2 k} \sum_{i=1}^k \Big(  \arccos (\sqrt{\Tilde{\lambda_i}})  -  \arccos( \sqrt{\lambda_i} ) \Big)^2 }  \leq \epsilon_p.
\label{eqn: property}
\end{align}

For simplicity, let $\Delta = \Tilde{p}_0 - p_0 $, $\arccos \sqrt{\Tilde{\lambda}_i} = \Tilde{x}_i$ and $\arccos \sqrt{\lambda_i} = x_i$. We have:
\begin{align}
    \Delta^2 &=  (\Tilde{p}_0 - p_0 )^2 \\
      &= \frac{4^2}{\pi^4 k^2} \Big( \sum_{i=1}^k ( \Tilde{x}_i^2 - x_i^2) \Big)^2.
\end{align}

Now we notice the following: for arbitrary two real numbers $x$ and $y$, there always exists a number $D$ such that $(x-y) \leq D(x+y)$. Apply this observation, and we have, for all $i$: $ \Tilde{x}_i^2 - x_i^2 = ( \Tilde{x}_i - x_i)(\Tilde{x}_i + x_i) \leq D_i (\Tilde{x}_i - x_i)^2 $. If we denote $D \equiv \max_i \{D_i\}_{i=1}^{k}$  then we have: 
\begin{align}
    \Delta^2 \leq \frac{4^2}{\pi^4 k^2} \Big( \sum_{i=1}^k  D_i(\Tilde{x}_i - x_i)^2     \Big)^2  \leq  \frac{4^2}{\pi^4 k^2} \Big( \sum_{i=1}^k  D(\Tilde{x}_i - x_i)^2     \Big)^2.
\end{align}
Taking the square root of both sides yields:
\begin{align}
    |\Delta|  &\leq D \frac{4}{\pi^2 k} \sum_{i=1}^k ( \Tilde{x}_i - x_i)^2,  \\
    &\leq D \epsilon_p = \mathcal{O}(\epsilon_p),
\end{align}
where the last line comes from Eqn. \ref{eqn: property}. Note that $|\Delta|$ is exactly $| \Tilde{p}_0 - p_0  | $, therefore, our proof is completed. 

\section{Block Encoding From Memory Model}
Here we explicitly show that the memory model \ref{lemma: datastructure} naturally encodes a  unitary block. We remark that it has been mentioned in the original work \cite{wossnig2018quantum} as well. \\

According to Lemma \ref{lemma: datastructure}, we have the following unitary:
\begin{align}
    U_M \ket{k}\ket{0} = \frac{1}{||A_i||} \sum_{i=1}^n A_{ik}\ket{k}\ket{i}, \\
    U_N \ket{0}\ket{j} = \frac{1}{||A||_F} \sum_{i=1}^m ||A_i|| \ket{i} \ket{j}.
\end{align}
We observe that:
\begin{align}
    \bra{0,j}U_N^\dagger U_M \ket{k}\ket{0} = \frac{A_{jk}}{|A|_F},
\end{align}
which matches perfectly with the definition of the block encoding \ref{def: blockencode}. Therefore, this is a very useful property of the memory model, making it easier to incorporate with quantum signal processing to construct a quantum algorithm, as outlined in the paper.

\end{document}